\documentclass[a4paper]{spie}  

\usepackage{amsmath,amsfonts,amssymb}
\usepackage{graphicx}
\usepackage[colorlinks=true, allcolors=blue]{hyperref}

\title{LEO-to-ground low elevation optical communication: optimization of an adaptive optics design robust to scintillation}

\author[a]{T. Vène}
\author[a]{A. Bonnefois-Montmerle}
\author[a]{L. M. Mugnier}
\author[a]{J.-M. Conan}
\affil[a]{DOTA, ONERA, Paris Saclay University, 92322 Châtillon, France}

\authorinfo{Further author information: (Send correspondence to Timothée Vène)\\Timothée Vène.: E-mail: timothee.vene@onera.fr}
 
\begin{document} 
\maketitle

\pagestyle{plain}    
\setcounter{page}{1} 

\begin{abstract}
To maximize the duration of optical downlinks with Low-Earth Orbit satellites, it is crucial to ensure the coupling into the single mode fiber of the ground terminal even at low elevations. Adaptive optics systems are designed to correct the wavefront deformation induced by atmospheric turbulence. 
However, at low elevations, amplitude fluctuations (or scintillation) challenge this correction. 
Here we propose a methodology to design a wavefront sensor that is robust to scintillation, taking the Shack-Hartmann WFS as an example. 
We present a slope estimator able to handle the large dynamic intensity range between subapertures. 
We then present an end-to-end simulation of the AO system to show the decrease in wavefront measurement error brought by a finer sampling of the pupil plane.
Finally, we discuss the feasibility of such design by means of a detailed AO error budget in the case of a LEO optical downlink.
\end{abstract}

\keywords{optical communication, adaptive optics, wavefront sensing, scintillation, Shack-Hartmann, strong perturbations}

\section{INTRODUCTION}
\label{sec:intro}  
In the case of Low Earth Orbit (LEO) satellites, there is a need for high-speed downlinks to transfer Earth observation images for instance.
To increase this throughput, optical solutions are increasingly being considered. 
Besides, the directivity of the laser beam ensures a better security compared to RF links. 
Nevertheless, atmospheric turbulence represents a major problem for optical communication. 
Indeed, because of the turbulence along the line of sight, the beam wavefront is distorted during propagation and this can lead to significant variations in the received signal and even interruptions due to fadings.
Hence the use of Adaptive Optics (AO) to compensate for this deformation and obtain a sufficiently high level of received signal coupled in a Single Mode Fiber (SMF) in the ground station terminal. 
However, the state-of-the-art AO system is designed to work in a certain turbulence regime: the weak perturbation regime which corresponds to the case of high elevations for a LEO satellite (between 90° and 30°). 
But at lower elevations, the part of the link taking place within the atmosphere becomes longer: not only do the phase deformations get more severe but also amplitude fluctuations (i.e scintillation) become stronger with a log-amplitude variance $\sigma_\chi^2$ exceeding 0.3. 
This is the strong perturbation regime where the performance of the AO system is not guaranteed. 
Nonetheless, these low elevations are crucial because LEO satellites spend half of the time they are in visibility of a given station between 30° and 10°. 

As pointed out by Primmerman \emph{et al.}(1995) \cite{primmerman1995atmospheric} and Barchers \emph{et al.}(2002) \cite{barchers2002evaluation}, scintillation affects an AO system by preventing the wavefront sensor (WFS) from correctly measuring the phase deformation inducing a biased correction of the wavefront by the deformable mirror.
In order to mitigate the effect of scintillation, alternative WFSs have been developed. 
For example, there are solutions based on self-referenced interferometry featuring a large dynamic range as presented in Rhoadarmer (2004)\cite{rhoadarmer2004development}, and in Notaras \emph{et al.}(2005)\cite{notaras2005direct}, and also some solutions using  variations of the Shack-Hartmann WFS (SHWFS) as in Lechner \emph{et al.}(2020) \cite{lechner2020adaptable}. 
However, at the moment, there is no theoretical study on wavefront sensing in the presence of scintillation describing the optimal WFS to use in this case.

In this communication, we illustrate the particularity of wavefront sensing in presence of scintillation by studying the SHWFS, a well-known WFS featuring a linear behavior and a wide sensing range while being simple to model. 
Scintillation causes fluctuations of intensity at two scales in the pupil plane of a SHWFS: wide intensity variations between subapertures, and non-uniform intensity distribution within a subaperture.
Concerning the latter, Mahé \emph{et al.}(2000) \cite{mahe2000scintillation} have shown that it induces a bias in the slope measurement performed by a SHWFS. 
Additionally, the intensity variations between subapertures require a slope estimator able to handle a large dynamic intensity range and saturated pixels.
We will address these two issues in order to optimize the design of the SHWFS in the presence of scintillation.

\section{Study of a new estimator} \label{section: JWLS}
To obtain orders of magnitude of the flux collected by the subapertures of a SHWFS in the strong perturbation regime, we simulate the  propagation through atmospheric turbulence as described in Vène \emph{et al.}(2023) \cite{vene2023} and using $C_n^2$ profiles from Osborn \emph{et al.}(2018) \cite{osborn2018optical}. 
We also take into account an AO design close to the design of the FEELINGS ground station~\cite{cyril2022feelings}: a telescope of $50\, \text{cm}$ diameter, and an AO system running at 5kHz, with a 25-by-25 subapertures SHWFS to which we assign 10nW on average. 
Due to scintillation, the intensity variations between subapertures span over 4 to 5 orders of magnitude, from $10^2$ to $10^6$ photons per subaperture. 
Thus, we need an estimator able to handle such a large dynamic intensity range. 
Furthermore, to improve the Signal-to-Noise Ratio (SNR) of the measurement, this estimator must be robust to saturated pixels.

The estimator that we developed is presented in Vène \emph{et al.}\cite{vene2025} and it is called the Joint Weighted Least Square (JWLS) estimator.
In this section, we compare our JWLS estimator to two well-studied slope algorithms: the center of gravity with a threshold at 10\% and the correlation. 
To compare their performance, we simulate  1000 occurrences of a spot moving at a subaperture focal plane, following a centered Gaussian distribution with $\sigma = 0.3$ pixel. 
For each spot, we apply a flux from $10^2$ to $10^7$ photons and  take into account both photon and detector noises. We take 30 photoelectrons of RON and $3.5\ 10 ^4$ photoelectrons as the sensor saturation level, both typical values for a state-of-the-art InGaAs camera.

   \begin{figure} [!htb]
   \begin{center}
   \begin{tabular}{c} 
   \includegraphics[scale=0.6]{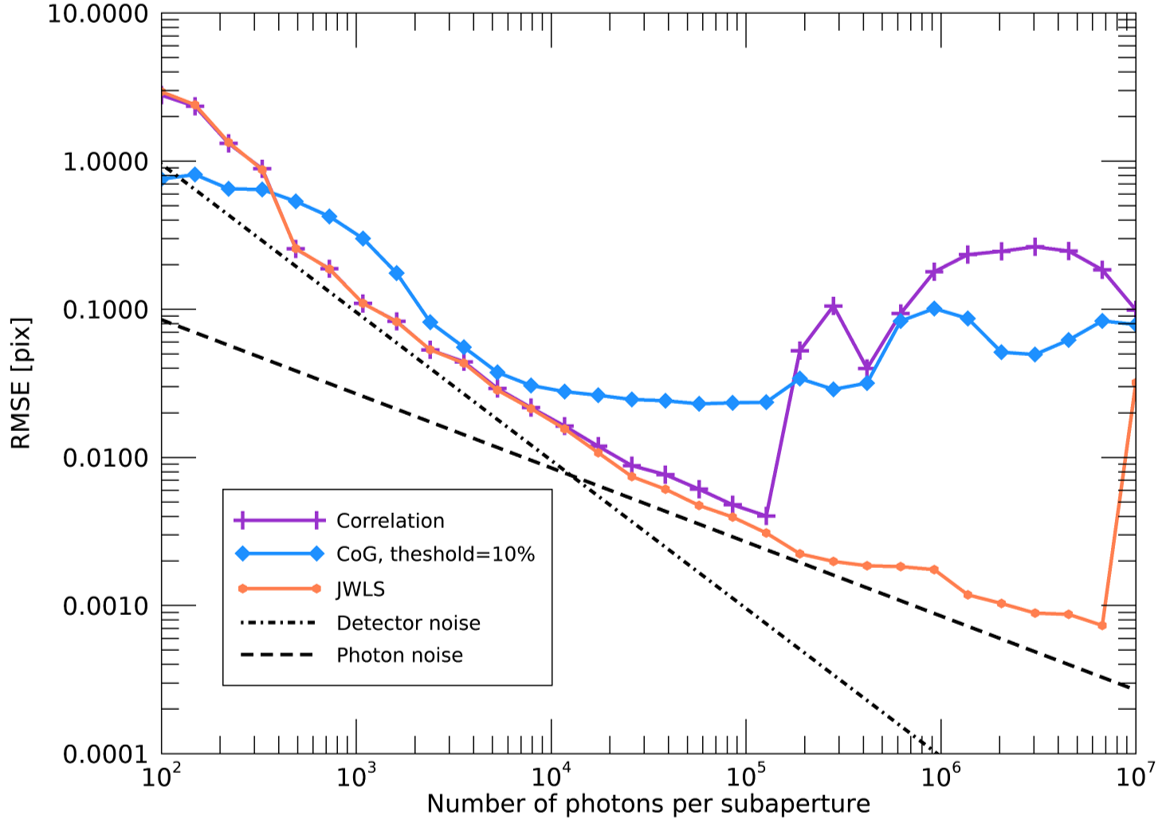}
   \end{tabular}
   \end{center}
   \caption[example] 
   { \label{fig:RMSE} 
Root Mean Square Error of slopes estimation as a function of the  number of photons in the image with sensor saturation at $3,5.10^4$ photoelectrons/pixel for: the correlation (purple line), the center of gravity algorithm (blue line) and the JWLS estimator (orange line). Both black dotted lines correspond to the correlation asymptote lines respectively in detector noise regime and in photon noise regime.}
   \end{figure} 
   
We obtain Figure~\ref{fig:RMSE}, showing the RMSE of each estimator as a function of the number of photons collected by the subaperture. The JWLS estimator is represented by the orange curve, the correlation by the purple curve and the thresholded center of gravity (TCoG) by the blue curve. 
We also show the asymptotes of the correlation error variance in detector noise and photon noise regimes from Thomas \emph{et al.}(2006) \cite{thomas2006}.  
In the detector noise regime (between $10^2$ and $10^4$ photons per subaperture), both the correlation and the JWLS estimator perform the same, which is expected since it can be shown that they are equivalent in this regime, as $\mathbf{w}$ is uniform.  
The TCoG performs better up to $4\ 10^2$ photons per subaperture thanks to the 10\%-threshold.
However, in a real implementation, these low SNR slope measurements would be considered not precise enough, and would be set to (0,0).
This would result in an RMSE of approximately $0.42\ \mathrm{rad}^2$, better than the RMSE of all considered estimators at signal levels below $4\ 10^2$ photons per subaperture.
In the photon noise regime (from $10^4$ to $1.4\ 10^5$ photons per subaperture), the TCoG error saturates as it suffers from non-linear effects due to the bias in its estimation; whereas the correlation follows its asymptote, and the JWLS outperforms slightly the correlation thanks to the weighting of the pixels. 
At fluxes higher than $1.4\ 10^5$ photons per subaperture, saturation affects both the correlation and the TCoG as we see that the error increases very significantly after this point. 
The JWLS estimator on the contrary, is accurate on approximately two more orders of magnitude of flux levels. 
This estimator is thus effectively more robust to sensor saturation.
It can be used when facing wide variations of flux between subapertures.
Besides, we can benefit from its robustness to sensor saturation, by increasing the assigned flux to the SHWFS: then, the number of low SNR subapertures decreases and the number of saturated subapertures increases while improving the overall measurement accuracy.

\section{Measurement error due to scintillation}
At the scale of a subaperture, intensity fluctuations at the pupil plane cause a measurement error as showed by Mahé \emph{et al.}(2000) \cite{mahe2000scintillation}.
It originates from the difference between the simplified and the physical reconstruction models of a SHWFS.
Indeed, assuming infinite field of view, continuous support and no noise, we have the following relation between the displacement of the spot and the phase gradient:

\begin{equation}
    \Delta x \propto \frac{\iint_S I(x,y)\frac{\partial \varphi (x,y)}{\partial x} \mathrm{d}x\mathrm{d}y}{\iint_S I(x,y)\mathrm{d}x\mathrm{d}y}
    \label{deltax1}
\end{equation}
where $\Delta x$ is the displacement of the spot along the x axis, $S$ is the support, $\varphi$  and $I$ are respectively the phase and the intensity of the incoming complex field.
It therefore corresponds to the average of the phase gradient weighted by the intensity distribution.

However, the reconstruction model assumes that the intensity in the subapertures is uniform. 
Eq.~\ref{deltax1} can then be simplified as: 
\begin{equation}
    \Delta x \propto \iint_S \frac{\partial \varphi (x,y)}{\partial x} \mathrm{d}x\mathrm{d}y
    \label{deltax2}
\end{equation}

This assumption is valid in the weak perturbation regime. 
But, considering strong perturbations and scintillation, the non-uniformity of the intensity distribution in the subapertures causes biases in the SHWFS measurements.
To our knowledge, the impact of scintillation on the measurements of a SHWFS has been studied at the scale of a subaperture only \cite{mahe2000scintillation}, but the quantification of the noise propagation into the reconstructed phase has not been addressed yet.

In this section, we study the effect of the SHWFS subapertures size on the reconstruction accuracy when facing scintillation.
We compare three SHWFS designs like in Vène \emph{et al.}(2022)\cite{vene2023revisiting}, but here with an end-to-end simulation of the AO system closed loop.
Considering a $48\, \text{cm}$ telescope, we study: a reference SHWFS featuring 24-by-24 subapertures (called SH24x24) whose size is approximately the average $r_0$ of the studied turbulence scenarios, $d_{subap}=2\, \text{cm}$; an over-sampled SHWFS featuring twice smaller subapertures (called SH48x48), $d_{subap}=1\, \text{cm}$; an under-sampled SHWFS featuring twice bigger subapertures (called SH12x12) to assess the benefit of aperture averaging, $d_{subap}=4\, \text{cm}$.
The slope measurement is performed by using a classical Center of Gravity algorithm with a threshold at 10\%.
The AO system is simulated to run at 10kHz.
Each SHWFS analyses the wavefront and then we reconstruct the phase on a given set of Zernike modes. 
The more subapertures the SHWFS design has, the more modes we can reconstruct. 
For the SH48x48, we reconstruct the first 40 Zernike radial orders; for the SH24x24, we reconstruct the first 21 Zernike radial orders; and for the SH12x12, we reconstruct the first 12 Zernike radial orders.

At each time $t$, we compare the reconstruction of each SHWFS affected by scintillation with the reconstruction performed by the same SHWFS if there was no scintillation, i.e respectively with the amplitude variations of the field or with a constant amplitude over the whole SHWFS pupil.
Besides, at this point the simulation takes into account neither noise nor sensor saturation in order to quantify only the error due to non-uniform intensity distribution within subapertures.

   \begin{figure} [!htb]
   \begin{center}
   \begin{tabular}{c} 
   \includegraphics[scale=0.6]{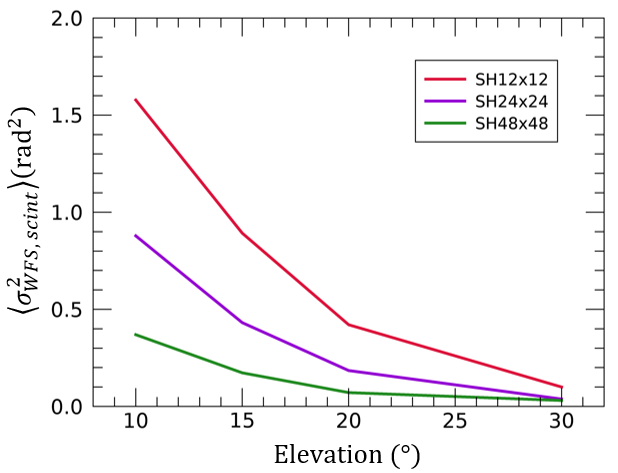}
   \end{tabular}
   \end{center}
   \caption[example] 
   { \label{fig:scintillation err} 
    Mean SHWFS error due to scintillation only as a function of elevation for 3 SHWFS: a SHWFS with 12-by-12 subapertures (red line), a SHWFS with 24-by-24 subapertures (purple line) and a SHWFS with 48-by-48 subapertures (green line).}
   \end{figure} 
   
Figure~\ref{fig:scintillation err} represents the measurement error due to scintillation averaged over time as a function of the elevation, in the case of the most severe atmospheric turbulence scenario featuring integrated parameters ranging from $r_0=2.6\mathrm{cm}$, $\sigma^2_\chi=0.13$ (at 30°) to $r_0=1.4\mathrm{cm}$, $\sigma^2_\chi=0.89$ (at 10°). 
The red curve corresponds to the SH12x12, the purple curve corresponds to the SH24x24, and the green curve corresponds to the SH48x48.
First, we observe that, for every SHWFS design, the error due to scintillation increases as the elevation decreases, as expected since the scintillation increases at lower elevations.
Besides, what is even more noticeable is that at every elevation, the smaller the subapertures the more precise the measurements.
Indeed, with smaller subapertures, the scintillation pattern are better sampled, and the intensity is more uniformly distributed within subapertures.
Thus, the model of Eq.~\ref{deltax2} is more valid in the case of the over-sampled SHWFS, and the measurements are thus more precise.

\section{SHWFS error term}
Considering an AO error budget, the global residual phase variance $\sigma^2_{\varphi_{res}}$ can be decomposed as the sum of four major contributions: 
\begin{equation}
    \sigma^2_{\varphi_{res}}=\sigma^2_{WFS}+\sigma^2_{fitting}+\sigma^2_{aliasing}+\sigma^2_{tempo} \ .
    \label{error budget}
\end{equation}
The last three error terms corresponding respectively to the fitting error, the aliasing error and the temporal error have already been studied in the literature \cite{1999aoa..book...91R}.
In the previous section, we quantify the part of the WFS error term depending solely on the scintillation.

Here, by using the same simulation as in the previous section and by taking into account noise and sensor saturation, we obtain the complete SHWFS error term.

\begin{figure} [!htb]
   \begin{center}
   \begin{tabular}{c} 
   \includegraphics[scale=0.55]{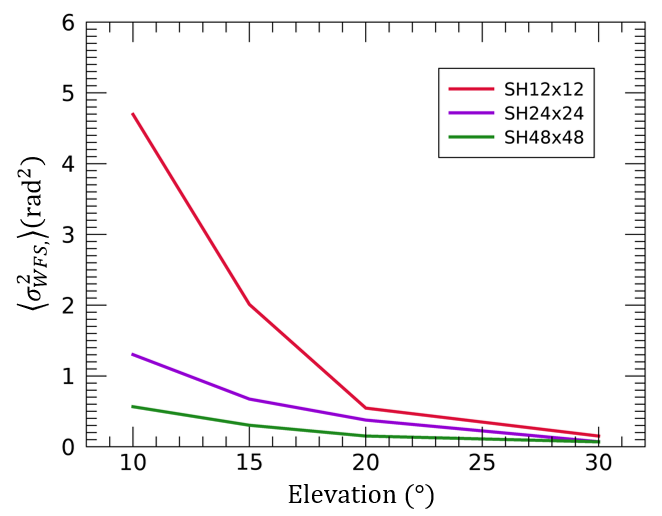}
   \end{tabular}
   \end{center}
   \caption[example] 
   { \label{fig:wfs err} 
    Mean measurement error of the SHWFS as a function of elevation for 3 SHWFS: a SHWFS with 12-by-12 subapertures (red line), a SHWFS with 24-by-24 subapertures (purple line) and a SHWFS with 48-by-48 subapertures (green line). On average, each SHWFS collects $1.7\ 10^5$ photons per subaperture and per frame. The saturation of the sensor occurs at $3.5\ 10^4$ photons per pixel.}
   \end{figure}

Figure~\ref{fig:wfs err} represents the error of the WFS as a function of the elevation with a sensor saturation occurring at $3.5\ 10^4$photoelectrons per pixel, a readout noise of 30 photoelectrons and a mean flux per subaperture of $1.7\ 10^5$ photons.
The flux per subaperture is the same for the three designs of SHWFS to compare their performance at equivalent SNR.
The red curve corresponds to the SH12x12, the purple curve corresponds to the SH24x24, and the green curve corresponds to the SH48x48.

We observe that as the elevation decreases, the error increases.
This is expected since at lower elevations both scintillation and wavefront distortions are more severe.

Regarding the AO error budget, a typical order of magnitude is $1\ \mathrm{rad}^2$ total.
Therefore, for this turbulence scenario, in order to maintain the optical link with a good AO correction at 10°, we must use the over-sampled SHWFS design.
Indeed, with this design, the WFS error term is low enough to keep a margin for the other error terms of $0.4\ \mathrm{rad}^2$.
Thus, the correction of a downlink with a LEO satellite at 10° is physically possible although it is really demanding in term of AO design.

Considering the SH48x48, a mean flux per subaperture of $1.7\ 10^5$ at 10kHz corresponds to an average power allocated to the SHWFS of $400\ \mathrm{nW}$ which is a significant portion of the link power, but is a conceivable trade-off to obtain a sufficient correction.

Furthermore, at $1.7\ 10^5$ photons per subapertures on average, and with a saturation level at $3.5\ 10^4$ photoelectrons per pixel, part of the subapertures are biased by the saturation in addition to the TCoG bias. 
We can thus expect that the WFS error term can be reduced by using the JWLS estimator presented in Section~\ref{section: JWLS}.

\section{Summary and perspectives}
In this paper, we showed that our JWLS estimator \cite{vene2025} is able to handle a wide dynamic intensity range and sensor saturation. 
This estimator enables us to reduce the number of low SNR subapertures and improve the overall measurement accuracy by allocating more flux to the SHWFS even though more subapertures may be saturated.
Then, we showed the effect of the size of a SHWFS subapertures on the robustness of its measurements in the presence of scintillation:  the subapertures must be small enough to sample correctly the scintillation patterns so that the intensity within a subaperture is uniform.
And finally, through an error budget, we discussed the feasibility of an AO system correcting a LEO downlink: it is achievable, although very demanding.

In future works, we will quantify the gain brought by the JWLS estimator regarding the slope estimation when facing turbulence scenarios that are representative of a LEO downlink at low elevations.

\acknowledgments 
The first author would like to thank ONERA and AID for funding his PhD.

\bibliography{report} 
\bibliographystyle{spiebib} 

\end{document}